\let\oldvec\vec
\documentclass[runningheads]{article}
\let\vec\oldvec

\usepackage{arxiv}
\usepackage{graphicx} 
\usepackage{booktabs} 
\usepackage{wrapfig} 
\usepackage{cite}
\usepackage[utf8]{inputenc} 
\usepackage[T1]{fontenc}    
\usepackage{xcolor}
\usepackage[bookmarks=false, unicode, colorlinks=true, urlcolor={blue!50!black}, linkcolor={red!50!black}, citecolor={green!50!black}]{hyperref}       
\usepackage{url}            
\usepackage{booktabs}       
\usepackage{nicefrac}       
\usepackage{microtype}      
\usepackage{lipsum}
\graphicspath{ {./ISVC19-VRVisionDeficit/} }
\usepackage{subfigure}
\usepackage{amsfonts, amsmath, amssymb} 
\usepackage{booktabs} 
\usepackage{wrapfig} 
\usepackage{sidecap}
\sidecaptionvpos{figure}{t}
\usepackage{indentfirst}
\usepackage[vlined,ruled]{algorithm2e}

\usepackage{array}
\newcolumntype{L}[1]{>{\raggedright\let\newline\\\arraybackslash\hspace{0pt}}p{#1}}
\newcolumntype{C}[1]{>{\centering\let\newline\\\arraybackslash\hspace{0pt}}p{#1}}
\newcolumntype{R}[1]{>{\raggedleft\let\newline\\\arraybackslash\hspace{0pt}}p{#1}}

\newcommand{\etal}{et al.~}

\usepackage{enumitem}
\usepackage{multirow}
\setlist[description]{leftmargin=8pt,labelindent=0pt,itemsep=0pt}
\setlist[itemize]{itemsep=0pt,parsep=0pt}
\setlist[enumerate]{itemsep=0pt,parsep=0pt, topsep=0pt}

\title{A Parametric Perceptual Deficit Modeling and Diagnostics Framework for Retina Damage using Mixed Reality}
\author{
  Prithul Aniruddha\\
    Department of Computer Science and Engineering\\
    University of Nevada, Reno\\
    Reno NV 89557
   \And
   Nasif Zaman\\
    Department of Computer Science and Engineering\\
    University of Nevada, Reno\\
    Reno NV 89557
   \And
 Alireza Tavakkoli \\
     Department of Computer Science and Engineering\\
    University of Nevada, Reno\\
    Reno NV 89557
 \And
 Stewart Zuckerbrod\\
    Houston Eye Associates, Inc.\\
    Houston TX 77401
}

\begin{document}

\maketitle              
\begin{abstract}
Age-related Macular Degeneration (AMD) is a progressive visual impairment affecting millions of individuals. Since there is no current treatment for the disease, the only means of improving the lives of individuals suffering from the disease is via assistive technologies. In this paper we propose a novel and effective methodology to accurately generate a parametric model for the perceptual deficit caused by the physiological deterioration of a patient's retina due to AMD. Based on the parameters of the model, a mechanism is developed to simulate the patient's perception as a result of the disease. This simulation can effectively deliver the perceptual impact and its progression to the patient's eye doctor. In addition, we propose a mixed-reality apparatus and interface to allow the patient recover functional vision and to compensate for the perceptual loss caused by the physiological damage. The results obtained by the proposed approach show the superiority of our framework over the state-of-the-art low-vision systems.
\end{abstract}

\section{Introduction}\label{sec:Intro}
There are several age-related eye diseases and conditions that drastically affect one's quality of life by causing permanent vision loss \cite{congdon2004causes}, \cite{massof1998systems}. These include: Age-related Macular Degeneration (AMD), Diabetic Eye Diseases, and Glaucoma, to name a few. According to the National Eye Institute an estimated 37 million adults in the U.S. over the age of 40 suffer from an age-related eye condition, such as, age-related Macular Degeneration (AMD), Glaucoma, Diabetic Retinopathy, and Cataracts \cite{NIE2018:EyeDes}. The leading condition affecting people over 40 is cataracts (24.4 million cases), followed by AMD (20 million cases), Diabetic Retinopathy (7.7 million cases), and Glaucoma (2.7 million cases). The total number of AMD, Diabetic Retinopathy, Cataract, and Glaucoma cases in the U.S. is expected to double between 2010 and 2050, according to the data from the National Eye Institute \cite{NIE2018:EyeStatsDouble}.

There is an effective surgical procedure to remove cataracts and treat patients. However, for other age-related conditions, such as AMD and other retinal diseases, there are no effective treatments to completely recover lost vision. Diabetic Retinopathy is the most common cause of vision loss in people with diabetes \cite{kahn1974blindness}, while advanced AMD is the leading cause of irreversible blindness and visual impairment in otherwise healthy individuals in the U.S. \cite{bressler2004age}. The blindness caused by AMD or Diabetic Retinopathy is due to the damage to the patient's retina that result in central vision loss \cite{evans2009quality}. However, the damage caused to the optic nerve as a result of Glaucoma produces gradual loss of peripheral vision in one or both eyes \cite{evans2009quality}. In this paper, our main focus is on central vision loss but the methodologies developed within this project can easily be expanded to cover peripheral vision loss.

\subsection{Motivations}\label{ssec:motivation}

As mentioned above, central vision loss is caused by the gradual deterioration of the center of the retina \cite{AMD2012}. Currently several tests, including, Humphrey Visual Field (HVF) Testing, Fluorescein Angiography, Visual Acuity Testing, and Optical Coherence Tomography (OCT), are administered by eye care professionals to determine the affected area on the patient's retina. While these tests could provide physicians with physiological impacts of the affected area within the eye, they fall short of providing a measurement of the perceptual impact on the patient's vision. Although the location and the physiological extent of the damage can be determined, the current assistive technologies such as NuEyes \cite{nuEyes}, eSight \cite{esightPaper}, \cite{eSightWeb}, and Vivid Vision \cite{VividVision2018}, do not utilize this information to provide site specific visual aid.

Current research indicates that there are a number of barriers that prevent the current technologies to provide each individual patient with specialized assistance in enhancing the remaining vision, \cite{kinateder2018using}. For example, there is a need to systematically examine how acuity level and visual field loss is associated with specific eye conditions, and how to utilize this connection to deliver specialized visual aid. There is also a need to utilize computer vision to enhance the provided visual aid to specific areas where the deficit in vision occurs. In this paper we address this gaps in the literature and develop a framework to significantly improve the current state of visual aid technologies for patients suffering from central vision loss.

\section{Literature Review}\label{sec:litRev}
\begin{figure}[tb]
\centering
\subfigure[]{\includegraphics[width=.45\linewidth]{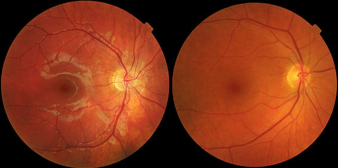}\label{fig:fundusAMD}}
\subfigure[]{\includegraphics[width=.45\linewidth]{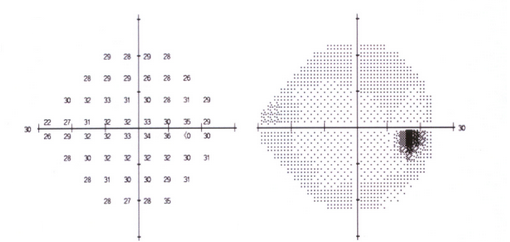}\label{fig:VF}}
\caption{(a) Retina Damage Caused by AMD (left), and Healthy Retina (right). (b) Visual Field Perimetry shows a map of the visual field and the areas within the visual field that may be affected.}
\end{figure}

Fundus photography is typically a great diagnostic tool for determining the physiological damage caused to the retina as a result of Age-related Macular Degeneration (AMD) -- Fig.~\ref{fig:fundusAMD}. In conjunction with visual field perimetry results (shown in Fig.~\ref{fig:VF}), physicians determine the physiological damage caused by the disease and track the progression of the condition.

\begin{figure}[htb]
  \centering
  \subfigure[]{\includegraphics[height=.9in]{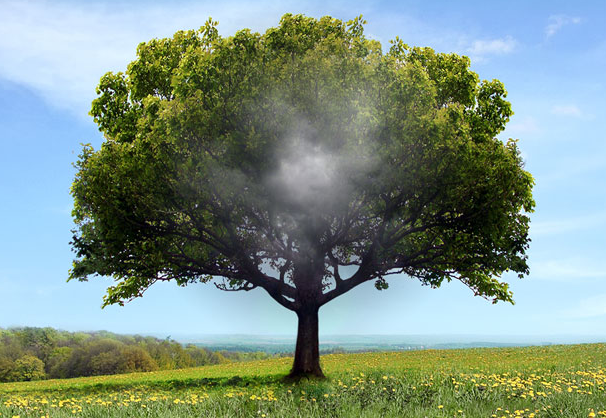}}
  \subfigure[]{\includegraphics[height=.9in]{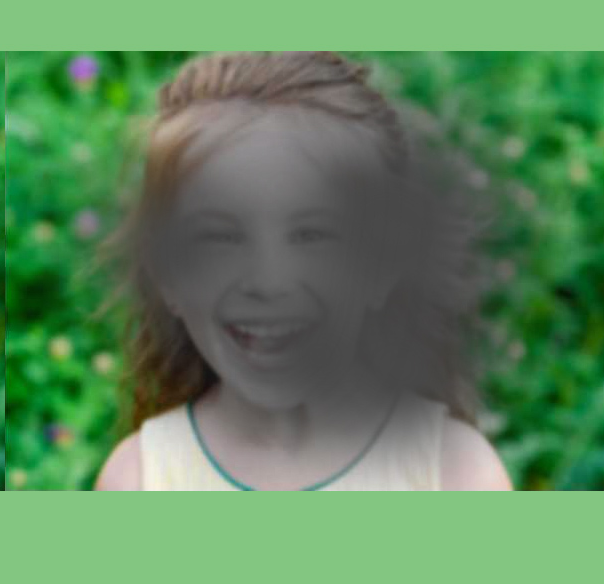}}
  \subfigure[]{\includegraphics[height=.9in]{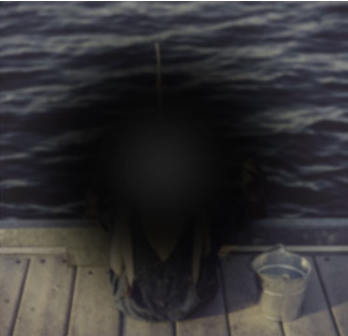}}
  \subfigure[]{\includegraphics[height=.9in]{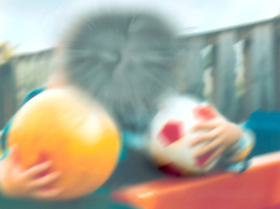}}
  \caption{\footnotesize{Different perceptual impacts of AMD on vision.}}\label{fig:AMD Visoin Perceptual Loss}
\end{figure}

The main problem is the difficulty in interpreting the perceptual impact of the physical damage. For example, AMD can cause a number of different perceptual effects in patient's vision, as shown in Fig.~\ref{fig:AMD Visoin Perceptual Loss}. Therefore, the main question to address in developing a model of perceptual deficit is, what does the patient see as a result of the physical damage to the retina, and can we correct it?

Many commercially available kits try to simulate various vision loss phenomenons~\cite{good-light, fork}. Usually they make use of goggles with easily changeable lenses to simulate different anomalies. Almutleb \etal used contact lenses to simulate central scotoma~\cite{central-scotoma}. Although these techniques, specially the ones that use goggles, are inexpensive, their setup is rather cumbersome with each disease requiring a different hardware setup. Moreover they cannot be modified anymore once built and therefore will lose their effectiveness as the disease progresses.

Thus, software-based simulation techniques could be employed to obtain more flexibility. In recent years, many software simulators have been developed. Lewis \etal developed a simulation inside Unreal Engine 3 to visualize six impairments including AMD~\cite{lewis2011simulating}. Currently a multitude of websites also provide interactive ways to simulate various impairments online, ~\cite{vision-loss-sim,vision-disorder,who}. However, these simulations work on a regular monitor. Moreover, they fall short of providing a complete binocular and stereoscopic simulation, thus are not an accurate and immersive representation of the visual loss.

Augmented Reality (AR) environments could be employed to deliver a more accurate simulation of the visual impairment. Vel{\'a}zquez \etal developed a simulator of several visual impairments for the normally sighted individuals~\cite{velazquez2016visual}. However, the system used camera feed from a single camera and therefore is not stereoscopic in nature. SimViz solved this problem by mounting two PlayStation eye cameras \cite{Ps4Eye} on an Oculus VR HMD to create a see-through AR display~\cite{ates2015immersive}.

Most of the above mentioned simulators are capable of visualizing central vision loss. However, none can accurately model the perceptual loss caused by actual physiological damage impacting an individual's retina. Therefore, although the state-of-the-art simulators may be used as tools, for example assisting in development of other applications with accessibility in mind~\cite{choo2017empath}, they cannot be employed to model a patient's perceptual vision loss nor are they capable of recovering functional vision caused by the disease.

Wearable AR HMD technology is still in it's infancy and most traditional approaches to low-vision, such as SimVis, deliver ad-hoc solutions. The Microsoft Hololens \cite{hololens} and the Magic Leap One \cite{MagicLeapOne} are among promising AR technologies on the horizon. However, while the potential is there, these devices are quite expensive and potentially out of reach for many patients \cite{robertson_2018}.

In order to alleviate the current roadblocks in the software and hardware technology for developing a unified infrastructure to allow for both simulation and vision recovery, we propose the computation of a parameterised model for the perceptual deficit generated by the patients themselves. In order to accomplish this, there needs to be an easy way to integrate various parameters of the model into an intuitive interface for patients, most of whom are advanced in age.

Current literature on interaction techniques in VR deals with 3D movement. Chatterjee \etal describe a hand gesture based interaction technique for desktop environments~\cite{chatterjee2015gaze+} and Pfeuffer \etal take a similar approach but introduce a VR environments~\cite{pfeuffer2017gaze+}. Both of these techniques, however, rely on eye gaze to focus on objects of interest. This can potentially cause problem for patients suffering from central vision loss, since fixation could be difficult for this population. Thus, in this paper, we focus on the development of a hand gesture based mechanism as a more appropriate design for our application.

\section{The Proposed Approach}
In this paper we utilize advances in the fields of Virtual Reality (VR) and Computer Vision (CV) in conjunction with the knowledge from current practices in the field of Ophthalmology to deliver transformative contributions to answer these questions. Specifically, this paper solves the problem of parameterizing the perceptual impairment and vision deficit in different AMD types based on the localized physiological damage. We will utilize the locus of the physiological damage to create parametric models for the visual deficit. This will lead to a VR central vision loss simulator for a variety of impairments associated with AMD.

\subsection{Modeling Perceptual Deficit} \label{sec:modelingAMD}
We propose a parametric model for the perceptual loss as a 4-tuple of the following form:
\begin{equation}\label{eq:Percp-Loss}
\mathcal{P}=(\mathbf{\Gamma},\mathbf{\Omega}_\lambda, \mathbf{R}_\theta, \mathbf\Psi)
\end{equation}

where $\mathbf{\Gamma}$ represents luminance degradation, $\mathbf\Omega_\lambda$ represents a parametrization of the visual field loss region with $\lambda$ as the cut-off value for the degradation determining the boundaries of $\mathbf\Omega_\lambda$, $\mathbf{R}_\theta$ is the rotational distortion matrix within $\mathbf\Omega_\lambda$, and $\mathbf\Psi$ is the a Sinusoidal mapping function representing the spatial distortion.

\subsubsection{Modeling Luminance Degradation Effects}\label{sssec:LuminanceChange}
We propose a Gaussian Mixture Model (GMM) \cite{cootes1999mixture} as a representative model for the degradation in luminance caused as a result of damage to the cone photoreceptors. Therefore, the proposed model for luminance degradation, $\mathbf\Gamma$, will be of the following form:

\begin{equation}\label{eq:luminEffect}
\mathbf{\Gamma}= \sum_{i=1}^{N}\omega_i\cdot\mathcal{N}_{\vec{\mu}_i,\sigma_i}(u,v)
\end{equation}
where $u$ and $v$ are the coordinate locations on the 2-D visual field, $N$ is the number of Gaussian kernels (Normal distributions) modeling the deficit in the luminance perception in the visual field, and $\omega_i$ is the amount of luminance deficit caused by each Gaussian kernel. Each Gaussian is represented by $\mathcal{N}_{\vec\mu_i,\sigma_i}(\cdot)$, where $\vec\mu_i=\left[\mu_i^u,\mu_i^v\right]^T$ represents the center and $\sigma$ represents the standard deviation of the distribution.

\begin{figure}[htb]
  \centering
  \subfigure[]{\includegraphics[height=1.1in]{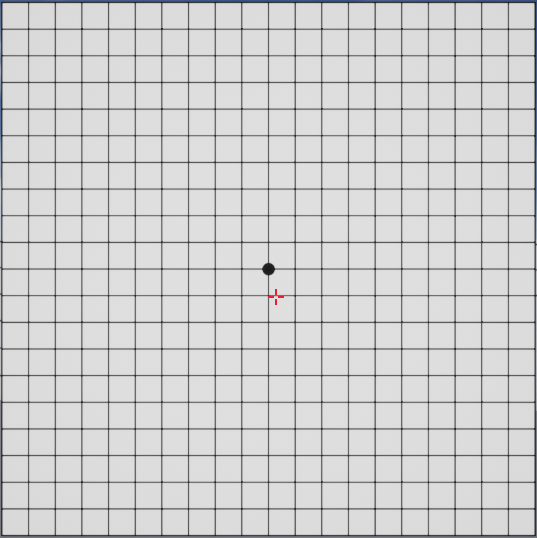}\label{fig:ilum-amsler}}
  \subfigure[]{\includegraphics[height=1.1in]{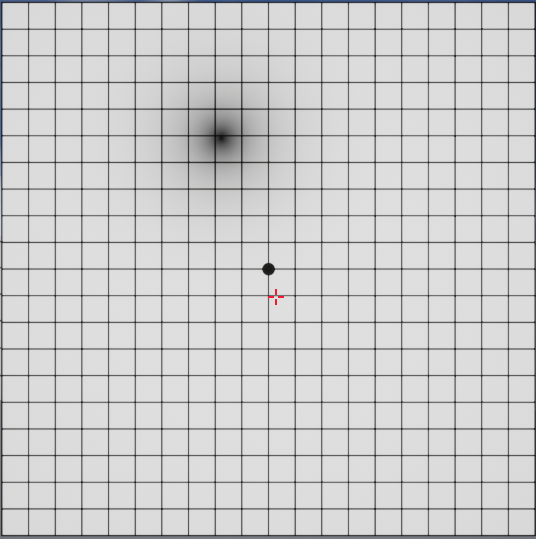}\label{fig:ilum-1G}}
  \subfigure[]{\includegraphics[height=1.1in]{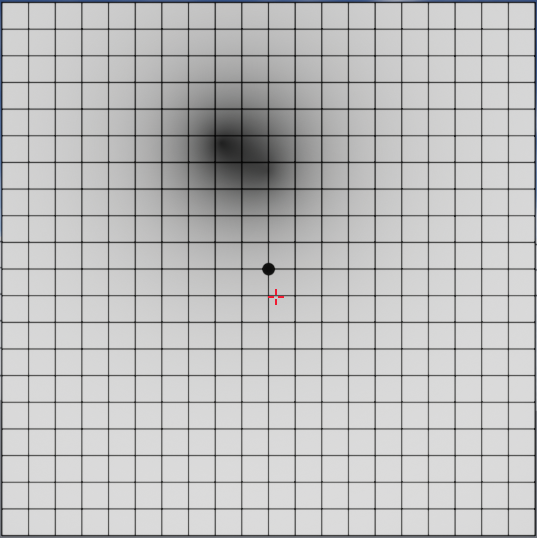}\label{fig:ilum-3G}}
  \subfigure[]{\includegraphics[height=1.1in]{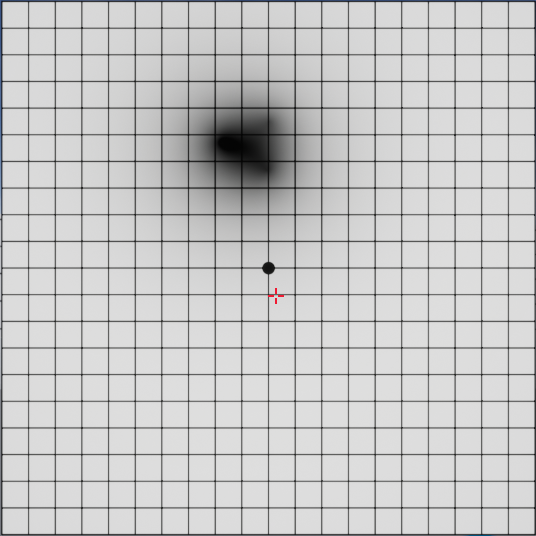}\label{fig:ilum-complex}}
  \caption{\footnotesize{Illumination Degradation $\mathbf\Gamma$, represented by the proposed parametric model.}}\label{fig:Illum-Model-Results}
\end{figure}

Fig.~\ref{fig:Illum-Model-Results} shows the results of the proposed illumination degradation model in affecting the vision on the Amsler grid (Fig. ~\ref{fig:ilum-amsler}). Fig.~\ref{fig:ilum-1G} shows the illumination degradation modeled by a single Gaussian.  A significant advantage of the proposed parametric model is in its ability to represent complex illumination degradations caused by the progressive retina damage. As shown in Fig.~\ref{fig:ilum-3G} and Fig.~\ref{fig:ilum-complex}, with the progression of the disease a complex mathematical formulation than a single Gaussian will be needed to model the degradation. Note that for each Gaussian kernel, the luminance deficit is the highest at the central location of that kernel ($\vec\mu$).

\subsubsection{Modeling Perceptual Deficit Region}\label{sssec:RegionDeficit}
\begin{figure}[htb]
  \centering
  \subfigure[]{\includegraphics[height=1.52in]{images/Lumin_deg_1Gn}\label{fig:ilum-complex2}}
  \subfigure[]{\includegraphics[height=1.52in]{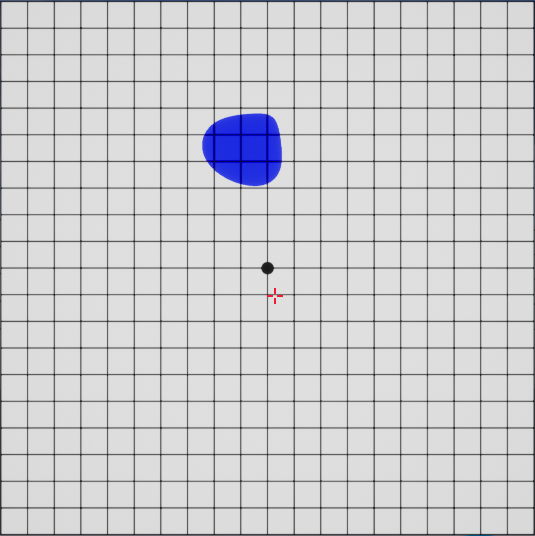}\label{fig:Region-complex}}
  \caption{\footnotesize{Perceptual deficit region $\Omega$, represented by the proposed parametric model. The area within the solid blue region represents illumination degradation of more than $\lambda$ percent.}}\label{fig:Region-Model-Results}
\end{figure}

Once the luminance degradation model is established in eq.(\ref{eq:luminEffect}), it will be easy to determine the region in the visual field in which the perceptual impact is significant (see Fig.~\ref{fig:Region-Model-Results}). Let's call this region $\mathbf\Omega$. Setting a cutoff value $0<\lambda<1$, the region $\mathbf\Omega$ can be determined as the following:
\begin{equation}\label{eq:region}
    \mathbf\Omega =\left\{(u,v)\in\mathbb R^2_{[0,1]} | \mathbf\Gamma(u,v)\leq\lambda\right\}
\end{equation}

\begin{figure}[htb]
  \centering
  \subfigure[]{\includegraphics[height=1.52in]{images/Lumin_deg_1Gn}\label{fig:ilum-complex3}}
  \subfigure[]{\includegraphics[height=1.52in]{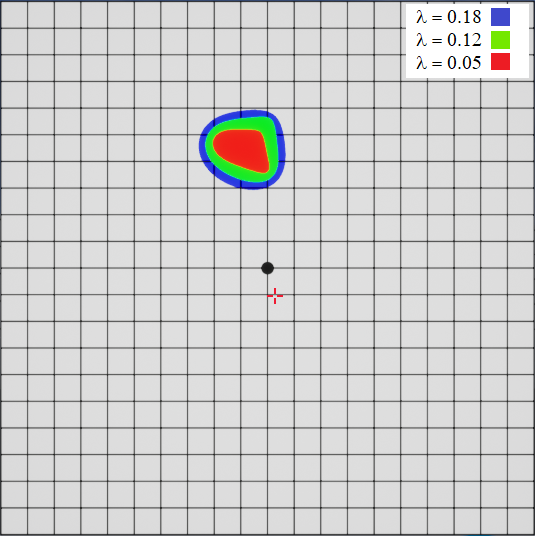}\label{fig:Region-Contours}}
  \caption{\footnotesize{Different perceptual deficit region $\Omega$ at various levels of degradation prominence $\lambda$. The area within the solid blue, green, and red regions represents illumination degradation of more than $\lambda = 18\%$, $\lambda = 12\%$, and $\lambda = 5\%$, respectively.}}\label{fig:Region-Model-Contours}
\end{figure}

The solid blue area in Fig.~\ref{fig:Region-complex} shows the perceptual deficit region $\mathbf\Omega$ for the modelled illumination degradation of Fig. ~\ref{fig:ilum-complex2}. Note that since $\lambda$ is a free parameter, it can control the boundary of the perceptual deficit region $\mathbf\Omega$. The larger the value of $\lambda$, the broader the regions $\mathbf\Omega$ will be. We can also visualize multiple regions with different prominent levels of illumination degradation (Fig.~\ref{fig:Region-Model-Contours}).

\subsubsection{Modeling Rotational Distortion}\label{sssec:rotation}
With the Loci of the perceptual damage determined by the cental positions ($\vec\mu_i$) of each Gaussian distribution in eq.(\ref{eq:luminEffect}), we can model the rotational distortion as a result of physiological damage. Recall from eq.(\ref{eq:Percp-Loss}) that rotational distortion, $\mathbf R_\theta$, is one of the components of the perceptual loss model, $\mathcal P$. Let $\theta$ be the angle of rotation, each point in the visual field will be rotated by the following rotation matrix:
\begin{equation}\label{eq:rotationEq}
    \mathbf{\hat{R}}_\theta = \begin{bmatrix}
                          \cos\theta & &-\sin\theta \\
                          \sin\theta & &\cos\theta \\
                        \end{bmatrix}
\end{equation}

However, since the perceptual impact decreases as we get farther away from the central location of each Gaussian kernel, the rotational distortion becomes less and less prominent. Therefore, we model the rotational distortion within the damaged region $\mathbf\Omega$ for each of the Gaussian kernels as:

\begin{equation}\label{eq:rot-perceptual-dist}
    \mathbf{R}_\theta=\sum_{i=1}^{N}\omega_i\mathcal{N}_{\vec\mu_i,\sigma_i}(u,v)\mathbf{\hat R}_{\theta_i}
\end{equation}

\begin{figure}[htb]
  \centering
  \subfigure[]{\includegraphics[height=1.12in]{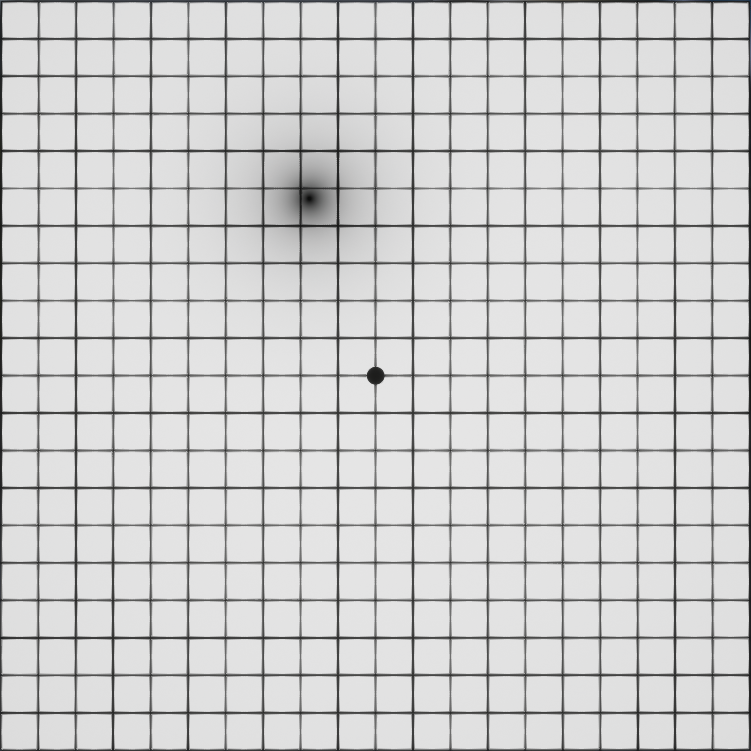}\label{fig:ilum-Decay-1G}}
  \subfigure[]{\includegraphics[height=1.12in]{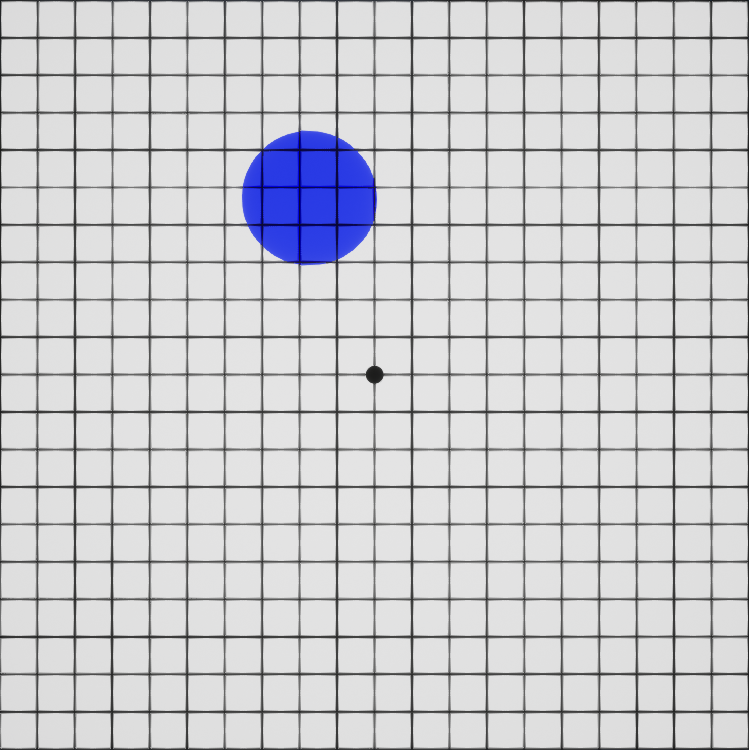}\label{fig:Rot-Region-1G}}
  \subfigure[]{\includegraphics[height=1.12in]{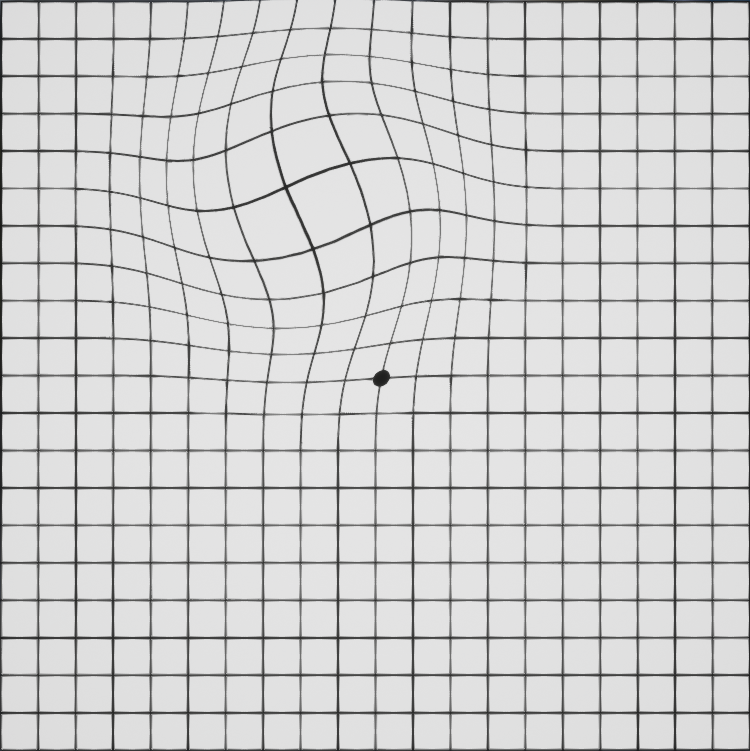}\label{fig:Rot-Distortion-1G}}
  \subfigure[]{\includegraphics[height=1.12in]{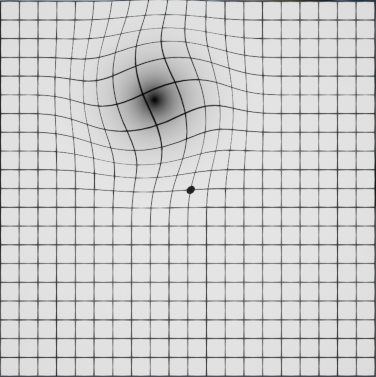}\label{fig:Rot-ilum-together-1G}}
  \caption{\footnotesize{Modeling rotational distortion: (a) A single Gaussian kernel illumination degradation. (b) The perceptual impact regions with $\lambda=0.5$. (c) the Rotational distortion with $\frac{\pi}{2}$ angle. (d) Both illumination degradation and rotational distortion.}}\label{fig:Rotational-Dist}
\end{figure}

The effects of this rotational distortion within the affected region of the visual field may be observed from Fig.~\ref{fig:Rotational-Dist}. In Fig.~\ref{fig:ilum-Decay-1G} the model utilizes a single Gaussian kernel with its impact region shown in Fig.~\ref{fig:Rot-Region-1G} at $\lambda=0.5$. A rotation of $\theta=\frac{\pi}{2}$ will cause the rotational distortion shown in Fig.~\ref{fig:Rot-Distortion-1G}. The effects of both the illumination degradation and rotational distortion can be observed in Fig.~\ref{fig:Rot-ilum-together-1G}.

\begin{figure}[hbt]
  \centering
  \subfigure[]{\includegraphics[height=1.5in]{images/Rot_Distandilum-1}\label{fig:Rot-ilum-together-1G1}}
  \subfigure[]{\includegraphics[height=1.5in]{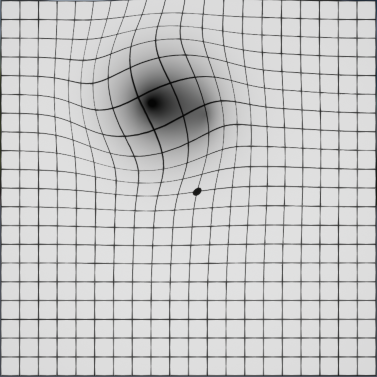}\label{fig:Rot-ilum-together-2G}}
  \subfigure[]{\includegraphics[height=1.5in]{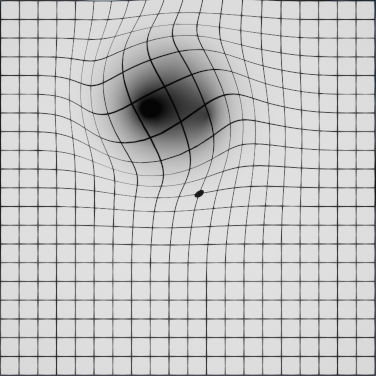}\label{fig:Rot-ilum-together-3G}}
  \caption{\footnotesize{Rotational distortion and illumination degradation as the disease progresses.}}\label{fig:Rotational-Dist-progressions}
\end{figure}

Fig.~\ref{fig:Rotational-Dist-progressions} shows the advantage of the proposed parameterized framework in modelling the progression of both the illumination degradation and rotational distortion as a result of the disease progression.

\subsubsection{Modeling Spatial Distortion}\label{sssec:spatialDist}
\begin{figure}[hbt]
  \centering
  \subfigure[]{\includegraphics[height=1.in]{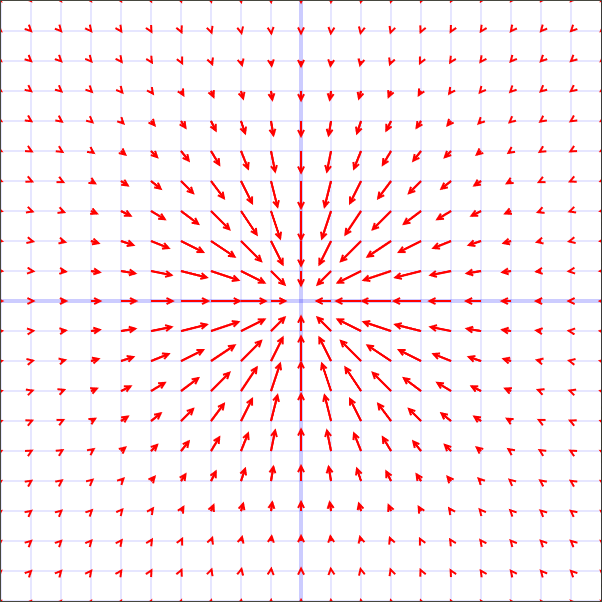}\label{fig:Spatial-VF-1G}}
  \subfigure[]{\includegraphics[height=1.in]{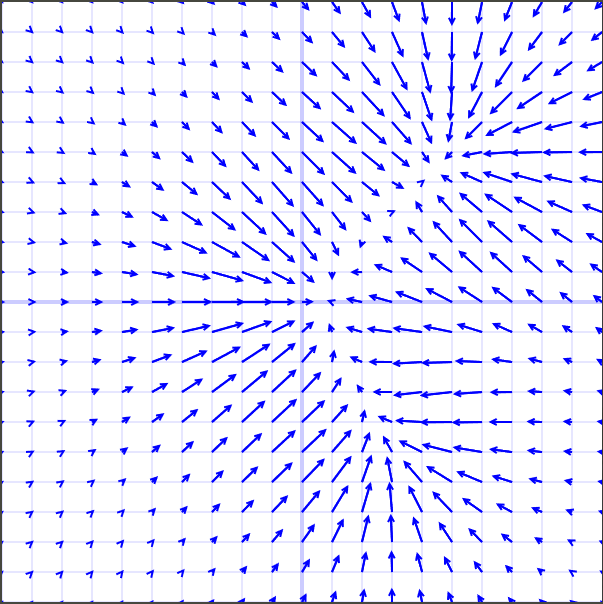}\label{fig:Spatial-VF-2G}}
  \caption{\footnotesize{Vector fields representing the spatial distortion maps, as modeled by: (a) a single Gaussian kernel, and (b) multiple Gaussian kernels.}}\label{fig:Spatial-VF}
\end{figure}

The final component of the perceptual deficit is the spatial distortion model $\mathbf\Psi$. This model represents the spatial shift perceived by the patient as a result of the damage to the retina that is not captured by the rotational distortion model described in section \ref{sssec:rotation}. This model is represented by a vector field dictating the spatial translation of points within the visual field. The complete spatial distortion vector field $\mathbf\Psi$ is defined as:

\begin{equation}\label{eq:VF-SpatialDist}
    \mathbf\Psi=\sum_{i=1}^{N}\mathcal{N}^i_{\vec{\mu_i},\sigma_i}(u,v)*\mathbf I_2*
                \begin{bmatrix}
                  \left(u-\mu_u^i\right) \\
                  \left(v-\mu_v^i\right) \\
                \end{bmatrix}
\end{equation}
where $\mathcal{N}^i$ represents each of the Gaussian deficit models with the mean of $\vec{\mu_i}$ and standard deviation of $\sigma_i$. $\mathbf{I_2}$ represents the $2\times2$ identity matrix, and $u$ and $v$ are coordinates within the visual field. To illustrate this spatial distortion effect, suppose we have a single scotoma at the central location of the visual field (i.e., $\left[ u\quad v \right]^T=\mathbf0$). The vector field representing the spatial distortion model will be of the following form (and shown in Fig.~\ref{fig:Spatial-VF-1G}):

\begin{equation}\label{eq:VF-SpatialDist1G}
    \mathbf\Psi\approx\begin{bmatrix}
                  e^{-\frac{(u-\mu_u)^2+(v-\mu_v)^2}{2\sigma^2}}*\left(u-\mu_u\right) \\
                  e^{-\frac{(u-\mu_u)^2+(v-\mu_v)^2}{2\sigma^2}}*\left(v-\mu_v\right) \\
                \end{bmatrix}
\end{equation}

Fig.~\ref{fig:Spatial-VF} shows the vector fields representing spatial transformations that the visual field undergoes as a result of the physiological damage. A single Gaussian kernel will generate a simple vector field shown in Fig.~\ref{fig:Spatial-VF-1G}, while a more complex spatial distortion will require a Gaussian Mixture Model as seen in Fig.~\ref{fig:Spatial-VF-2G}.

\begin{figure}[hbt]
  \centering
  \subfigure[]{\includegraphics[height=1.in]{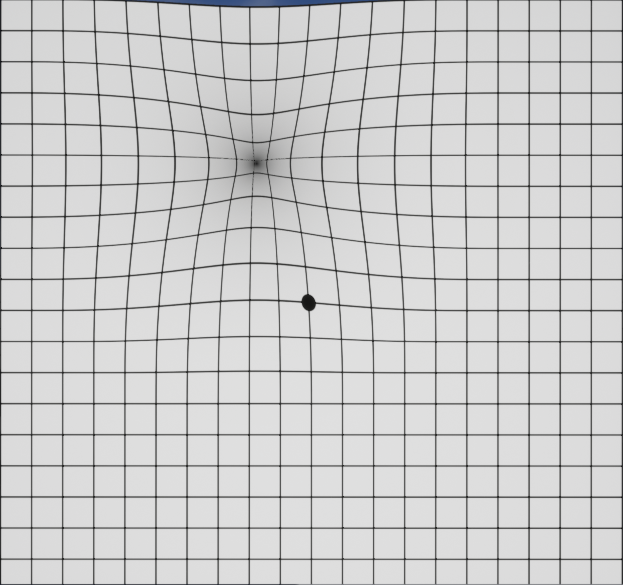}\label{fig:Spatial-dist-1G}}
  \subfigure[]{\includegraphics[height=1.in]{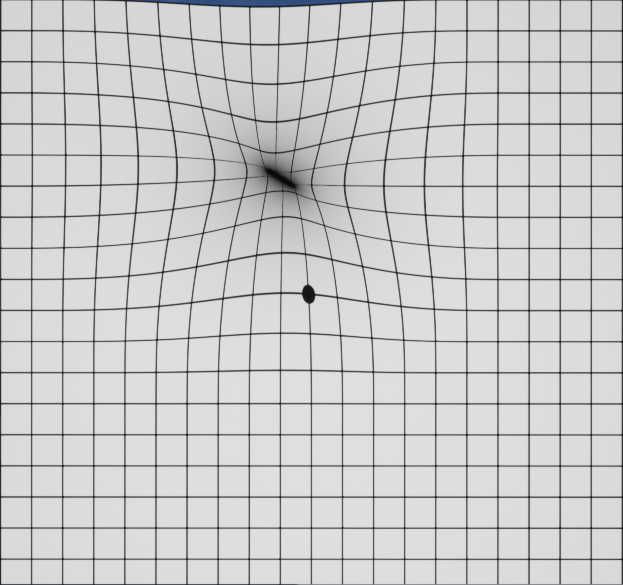}\label{fig:Spatial-dist-2G}}
  \subfigure[]{\includegraphics[height=1.in]{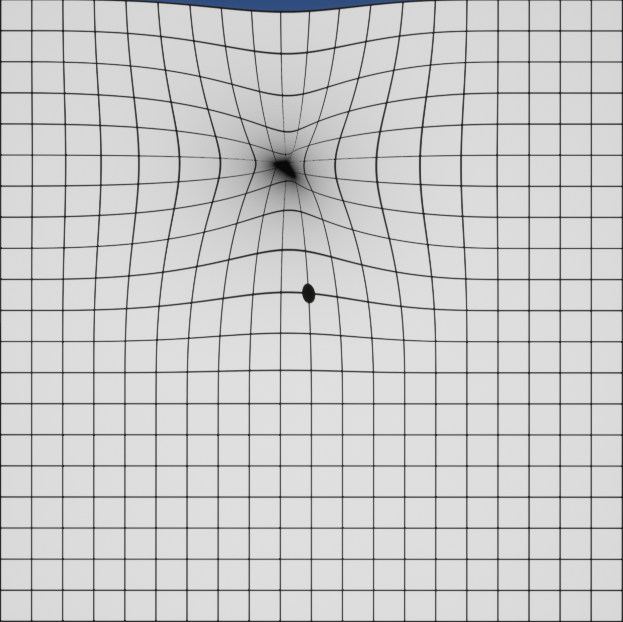}\label{fig:Spatial-dist-3G}}
  \caption{\footnotesize{Spatial distortion of the Amsler grid as represented by the $\mathbf\Psi$ component of the proposed model. Note the model adaptation and its ability to capture disease progression: modeling spatial distortion with (a) a single Gaussian kernel, (b) two Gaussian kernels, and (c) three Gaussian kernels.}}\label{fig:Spatial-dist}
\end{figure}

The strength of the proposed model for representing the spatial distortions caused by the disease can be seen from Fig.~\ref{fig:Spatial-dist}. At the early stages of the disease a single Gaussian kernel may be sufficient to model the distortions (Fig.~\ref{fig:Spatial-dist-1G}), but as the disease progresses more complex models will be required. The proposed mixture model shows the flexibility to represent the distortion changes as the disease progresses without the need to fundamentally change the model, but rather increase the number of Gaussian kernels (Fig.~\ref{fig:Spatial-dist-2G}-Fig.~\ref{fig:Spatial-dist-3G}).

\begin{figure}[ht]
\centering
\subfigure[]{
   \includegraphics[width=.48\linewidth]{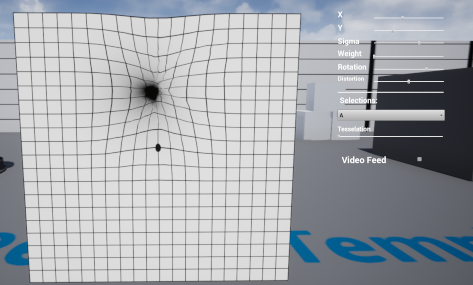}\label{fig:Res-AmslerMapUI}}
\subfigure[]{\includegraphics[width=.48\linewidth]{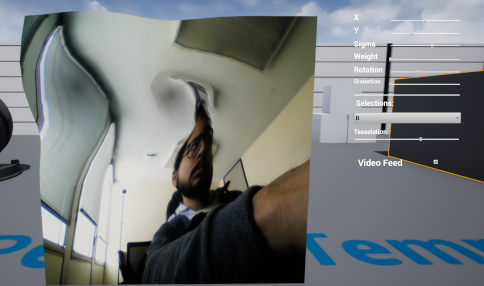}\label{fig:res-MRVideo}}
   \caption{(a) The diagnostic interface allowing the patient to establish the perceptual deficit model and populate its parameters in the Diagnostic VR environment. (b) The functional vision recovery allowing the video feed from a webcam corrected by the patient's perceptual deficit model and mapped to patient's eye.}
\end{figure}

    \begin{figure}[ht]
    \centering
    \includegraphics[width=.85\linewidth]{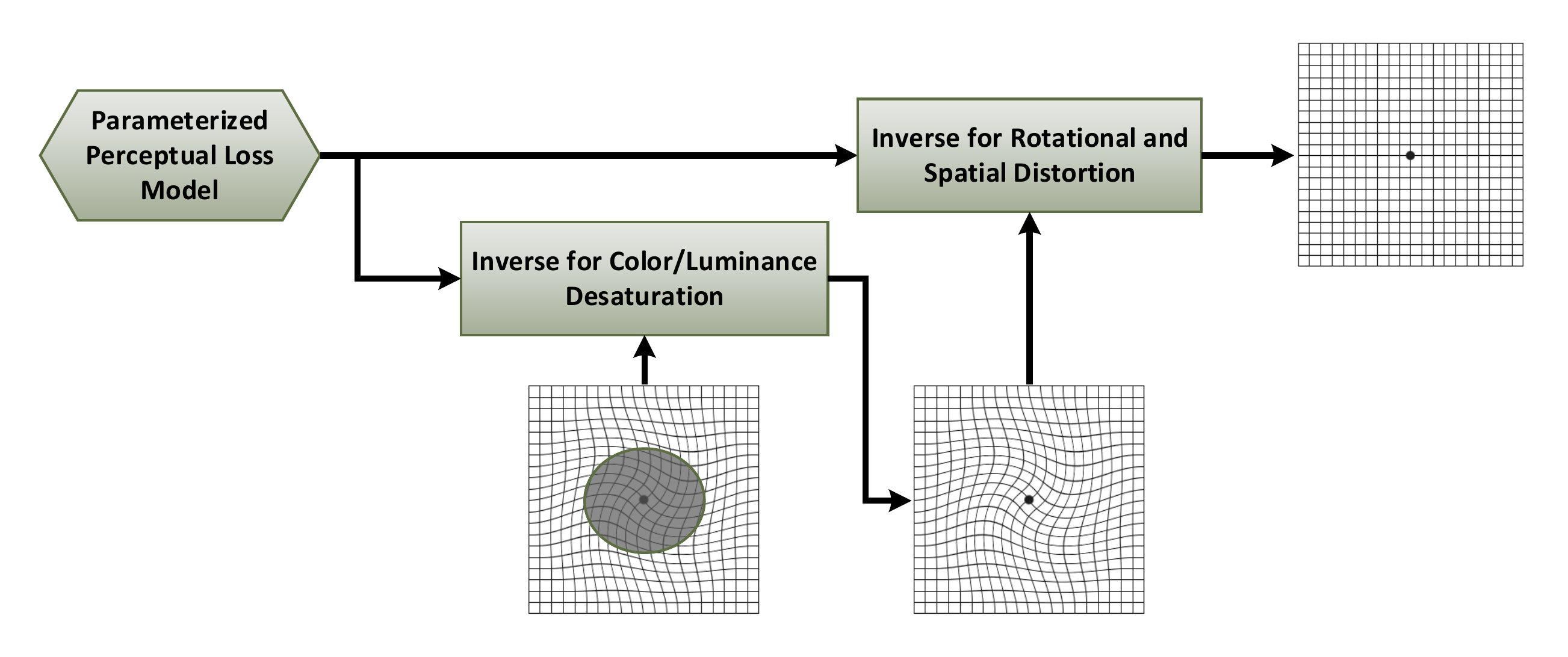}
    \caption{\footnotesize{The dichoptic solution for vision compensation by enhancing the remaining vision in the affected eye determines the vision loss region ($W$) and the parameters of the perceptual loss from Model $\mathcal{P}$.
    }}\label{fig:Enhancement}
    \end{figure}

\section{Experimental Results}\label{sec:Results}

The proposed framework is a comprehensive diagnostic and vision recover system. Therefore, the patients will utilize the system in two phases: The VR Diagnostic Mode, and the AR Vision Compensation Mode.

In the first mode (shown in Fig.~\ref{fig:Res-AmslerMapUI}) the Amsler grid is shown to the patient in a neutral VR environment with the proposed user interfaces to allow the patient to interact with parameters of their perceptual deficit model. Changing the parameters updates the material being rendered for the patient's eyes in real-time. Our goal here is to have the user closely mimic what they see with their affected eye in order to populate the parameters of their perceptual deficit model in the affected eye.

Once the the patient accurately model their perceptual deficit, the remainder of the operation for recovering functional vision will take place in the AR mode. In this mode the VR HMD works as an AR glass, by presenting the patient with the live video of the environment captured through the stereoscopic video cameras. The images in the video feed shown to the unaffected eye is, however, distorted by with the perceptual loss model that the user came up in the Amsler grid mode. This gives the user an opportunity to test how the distortion looks in real life. Fig.~\ref{fig:res-MRVideo} shows the uncorrected vision of a patient as observed by a physician. The inverse of the parametric model is applied and the patient sees the environment as if their eye were not affected.

Based on the model generated by the proposed framework in the VR Diagnostics mode, an inverse function for the perceptual vision loss is calculated. This inverse model is then applied in the AR mode to the live-stream videos recorded and rendered to each individual eye. The process is shown in Fig.~\ref{fig:Enhancement}. As seen in the figure, the inverse function for the parametric vision loss will compensate for the perceptual impact of the vision loss to recover as much functional vision as possible.

\section{Conclusions and Future Work}
In this paper we presented a parametric model for the perceptual deficit caused by Age-related Macular Degeneration (AMD). This model bridges the gap between our current state of knowledge about the physiological damage caused to the retina and its perceptual effect on the patient's vision. We developed a mixed-reality system that allows the patient to configure and update the parameters of their perceptual deficit in a VR mode. Once the patient's unique vision loss is modeled, the system is employed in AR mode to perform the inverse model on the affected eye to compensate for the perceptual deficit and recover functional vision. The experiments show that the proposed system captures the patient's vision loss and allows for significant recovery of the functional vision.

One future direction to this work includes modeling other vision loss and visual anomalies such as Glaucoma, Diabetic Retinopathy, etc. Our team is also conducting human trails in order to validate the usability, efficacy, and acceptability of the proposed system.

\bibliographystyle{unsrt}
\bibliography{ISVC19-VRVisionDeficit/references}

\begin{thebibliography}{10}

\bibitem{congdon2004causes}
Nathan Congdon, Benita O'Colmain, CC~Klaver, Ronald Klein, Beatriz Mu{\~n}oz,
  David~S Friedman, John Kempen, Hugh~R Taylor, and Paul Mitchell.
\newblock Causes and prevalence of visual impairment among adults in the united
  states.
\newblock {\em Archives of Ophthalmology (Chicago, Ill.: 1960)},
  122(4):477--485, 2004.

\bibitem{massof1998systems}
Robert~W Massof.
\newblock A systems model for low vision rehabilitation. ii. measurement of
  vision disabilities.
\newblock {\em Optometry and vision science: official publication of the
  American Academy of Optometry}, 75(5):349--373, 1998.

\bibitem{NIE2018:EyeDes}
{National Eye Institute}.
\newblock Prevalence of adult vision impairment and age-related eye diseases in
  america.
\newblock \url{https://nei.nih.gov/eyedata/adultvision_usa}.
\newblock Accessed: 9-15-2018.

\bibitem{NIE2018:EyeStatsDouble}
{National Eye Institute}.
\newblock Statistics and data: The most common eye diseases: Nei looks ahead.
\newblock \url{https://nei.nih.gov/eyedata}.
\newblock Accessed: 9-15-2018.

\bibitem{kahn1974blindness}
Harold~A Kahn and Rita Hiller.
\newblock Blindness caused by diabetic retinopathy.
\newblock {\em American journal of ophthalmology}, 78(1):58--67, 1974.

\bibitem{bressler2004age}
Neil~M Bressler.
\newblock Age-related macular degeneration is the leading cause of blindness...
\newblock {\em Jama}, 291(15):1900--1901, 2004.

\bibitem{evans2009quality}
Keith Evans, Simon~K Law, John Walt, Patricia Buchholz, and Jan Hansen.
\newblock The quality of life impact of peripheral versus central vision loss
  with a focus on glaucoma versus age-related macular degeneration.
\newblock {\em Clinical ophthalmology (Auckland, NZ)}, 3:433, 2009.

\bibitem{AMD2012}
Rama~D. Jager, William~F. Mieler, and Joan~W. Miller.
\newblock Age-related macular degeneration.
\newblock {\em New England Journal of Medicine}, 358(24):2606--2617, 2008.
\newblock PMID: 18550876.

\bibitem{nuEyes}
{NuEyes Inc.}
\newblock {NuEyes}.
\newblock \url{https://nueyes.com/nueyes-pro/}.
\newblock Accessed: 2018.

\bibitem{esightPaper}
Robert Devenyi.
\newblock Wearable technology expands mobility for visually impaired.
\newblock {\em Ophthalmology Times}, 2016.

\bibitem{eSightWeb}
e{Sight}.
\newblock e{Sight}.
\newblock \url{https://www.esighteyewear.com/homex}.
\newblock Accessed: 2018.

\bibitem{VividVision2018}
{Vivid Vision}.
\newblock {Vivid Vision}.
\newblock \url{https://www.seevividly.com/}.
\newblock Accessed: 2018-09-30.

\bibitem{kinateder2018using}
Max Kinateder, Justin Gualtieri, Matt~J Dunn, Wojciech Jarosz, Xing-Dong Yang,
  and Emily~A Cooper.
\newblock Using an augmented reality device as a distance-based vision
  aid—promise and limitations.
\newblock {\em Optometry and Vision Science}, 95(9):727, 2018.

\bibitem{good-light}
{Good-lite}.
\newblock Visualeyes vision simulator glasses.
\newblock \url{https://www.good-lite.com/Details.cfm?ProdID=766}.
\newblock Accessed: 2019-03-24.

\bibitem{fork}
{Fork in the Road Website}.
\newblock Fork in the road : Low vision simulators.
\newblock
  \url{https://www.lowvisionsimulators.com/collections/find-the-right-low-vision-simulator/}.
\newblock Accessed: 2019-03-24.

\bibitem{central-scotoma}
ES~Almutleb, A~Bradley, J~Jedlicka, and SE~Hassan.
\newblock Simulation of a central scotoma using contact lenses with an opaque
  centre.
\newblock {\em Ophthalmic Physiol Opt 2018}, 38:76--87, 2017.

\bibitem{lewis2011simulating}
James Lewis, David Brown, Wayne Cranton, and Robert Mason.
\newblock Simulating visual impairments using the unreal engine 3 game engine.
\newblock In {\em 2011 IEEE 1st International Conference on Serious Games and
  Applications for Health (SeGAH)}, pages 1--8. IEEE, 2011.

\bibitem{vision-loss-sim}
{}Vision~Loss Simulator.
\newblock Vision loss simulator.

\bibitem{vision-disorder}
Vision disorders.
\newblock
  \url{https://www.genworth.com/aging-and-you/health/vision-disorders.html#/}.
\newblock Accessed: 2019-03-22.

\bibitem{who}
{World Health Organization}.
\newblock Blindness and vision impairment.
\newblock
  \url{https://www.who.int/news-room/fact-sheets/detail/blindness-and-visual-impairment}.
\newblock Accessed: 2019-03-22.

\bibitem{velazquez2016visual}
Ramiro Vel{\'a}zquez, Claudia~N S{\'a}nchez, and Edwige~E Pissaloux.
\newblock {Visual impairment simulator based on the Hadamard product}.
\newblock {\em Electronic Notes in Theoretical Computer Science}, 329:169--179,
  2016.

\bibitem{Ps4Eye}
Sarah Stocker.
\newblock Playstation eye, a little more info, 2007.
\newblock
  \url{https://blog.us.playstation.com/2007/10/10/playstation-eye-a-little-more-info/}.

\bibitem{ates2015immersive}
Halim~Cagri Ates, Alexander Fiannaca, and Eelke Folmer.
\newblock Immersive simulation of visual impairments using a wearable
  see-through display.
\newblock In {\em Proceedings of the Ninth International Conference on
  Tangible, Embedded, and Embodied Interaction}, pages 225--228. ACM, 2015.

\bibitem{choo2017empath}
Kenny Tsu~Wei Choo, Rajesh~Krishna Balan, Tan~Kiat Wee, Jagmohan Chauhan,
  Archan Misra, and Youngki Lee.
\newblock Empath-d: Empathetic design for accessibility.
\newblock In {\em Proceedings of the 18th International Workshop on Mobile
  Computing Systems and Applications}, pages 55--60. ACM, 2017.

\bibitem{hololens}
Microsoft hololens mixed reality technology for business.
\newblock \url={https://www.microsoft.com/en-us/hololens}.
\newblock Accessed: 2019-05-13.

\bibitem{MagicLeapOne}
{Magic Leap Inc.}
\newblock Magic leap one, 2019.
\newblock \url{https://www.magicleap.com/magic-leap-one}.

\bibitem{robertson_2018}
Adi Robertson.
\newblock I tried magic leap and saw a flawed glimpse of mixed reality's
  amazing potential, Aug 2018.

\bibitem{chatterjee2015gaze+}
Ishan Chatterjee, Robert Xiao, and Chris Harrison.
\newblock Gaze+ gesture: Expressive, precise and targeted free-space
  interactions.
\newblock In {\em Proceedings of the 2015 ACM on International Conference on
  Multimodal Interaction}, pages 131--138. ACM, 2015.

\bibitem{pfeuffer2017gaze+}
Ken Pfeuffer, Benedikt Mayer, Diako Mardanbegi, and Hans Gellersen.
\newblock Gaze+ pinch interaction in virtual reality.
\newblock In {\em Proceedings of the 5th Symposium on Spatial User
  Interaction}, pages 99--108. ACM, 2017.

\bibitem{cootes1999mixture}
Timothy~F Cootes and Christopher~J Taylor.
\newblock {A mixture model for representing shape variation}.
\newblock {\em Image and Vision Computing}, 17(8):567--573, 1999.

\end{thebibliography}

\end{document}